\documentclass[useAMS,usenatbib]{mn2e}
\usepackage{times,epsfig,url}

\title[The X-ray properties of young radio-loud AGN]
{The X-ray properties of young radio-loud AGN}

\author[J. Vink, I. Snellen, K-H. Mack, R. Schilizzi]{Jacco Vink$^{1,2}$\thanks{E-mail: j.vink@astro.uu.nl}, Ignas Snellen$^{3}$,
Karl-Heinz Mack$^{4}$ and Richard Schilizzi$^{5}$\\
$^{1}$Astronomical Institute, University Utrecht, P.O. Box 80000, NL-3508TA Utrecht, The Netherlands\\
$^{2}$SRON Netherlands Institute for Space Research, Sorbonnelaan 2, NL-3584CA Utrecht, The Netherlands\\
$^{3}$Leiden Observatory, Niels Bohrweg 2, NL-2300RA Leiden, The Netherlands\\
$^{4}$INAF - Istituto di Radioastronomia, Via Gobetti 101, I-40129 Bologna, Italy \\
$^{5}$International SKA project office,  P.O. Box 2, NL-7990AA Dwingeloo, The Netherlands\\
}
\begin{document}

\date{}

\pagerange{\pageref{firstpage}--\pageref{lastpage}} \pubyear{2002}

\maketitle

\label{firstpage}

\newcommand{\apj}{{ApJ}}
\newcommand{\apjs}{{ApJS}}
\newcommand{\apjl}{{ApJ}}
\newcommand{\aj}{{AJ}}
\newcommand{\aap}{{A\&A}}
\newcommand{\aaps}{{A\&AS}}
\newcommand{\nat}{{Nat}}
\newcommand{\jetp}{{JETP}}
\newcommand{\mnras}{{MNRAS}}
\newcommand{\phrvl}{{PhRvL}}
\newcommand{\phrc}{{PhRvC}}
\newcommand{\prc}{{PhRvC}}
\newcommand{\araa}{{ARA\&A}}
\newcommand{\pasj}{{PASJ}}
\newcommand{\pasp}{{PASP}}
\newcommand{\npa}{{NuPhA}}
\newcommand{\iaucirc}{{IAU circ.}} 
\newcommand{\aplett}{{Astrophysical Letters}} 

\newcommand{\rvmp}{{\it Rev. Mod. Physics}}
\newcommand{\xmm}{{\it XMM-Newton}}
\newcommand{\chandra}{{\it Chandra}}
\newcommand{\asca}{{\it ASCA}}
\newcommand{\rosat}{{\it ROSAT}}
\newcommand{\einstein}{{\it Einstein}}
\newcommand{\cangeroo}{{\it CANGEROO}}
\newcommand{\whipple}{{\it Whipple}}
\newcommand{\hegra}{{\it HEGRA}}
\newcommand{\hess}{{\it HESS}}
\newcommand{\smm}{{\it SMM}}
\newcommand{\sax}{{\it BeppoSAX}}
\newcommand{\rxte}{{\it RXTE}}
\newcommand{\osse}{{\it OSSE}}
\newcommand{\egret}{{\it CGRO-EGRET}}
\newcommand{\integr}{{\it INTEGRAL}}
\newcommand{\glast}{{\it GLAST}}
\newcommand{\comptel}{{\it COMPTEL}}
\newcommand{\cgro}{{\it CGRO}}

\newcommand{\xspec}{{\it xspec}}

\newcommand{\oiii}{\hbox{[O\,III]}}

\newcommand{\msun}{{$M_{\odot}$}}
\newcommand{\nh}{{$N_{\rm H}$}}
\newcommand{\ep}{{e$^+$e$^-$}}
\newcommand{\fluxunit}{{ph\,cm$^{-2}$s$^{-1}$}}
\newcommand{\kms}{{km\,s$^{-1}$}}
\newcommand{\ndot}{{\dot{N}_{UV}}}
\newcommand{\NH}{{N_{\rm H}}}
\newcommand{\loiii}{{L_{\rm [O III]}}}

\begin{abstract}
We present \xmm\ observations of a complete sample of 
five archetypal young radio-loud AGN, also known as Compact Symmetric Objects 
(CSO) or Gigahertz Peaked Spectrum (GPS) sources. They are among the brightest
and best studied GPS/CSO sources in the sky, with radio powers in the range
L$_{\rm{5GHz}}$=10$^{43-44}$ erg s$^{-1}$ and with four sources having
measured kinematic ages of 570 to 3000 years. 
All five sources are detected, 
and have 2-10 keV luminosities ranging
from 0.5 to 4.8$\times 10^{44}$ erg s$^{-1}$. 
A detailed analysis was performed, comparing the X-ray luminosities and 
$\NH$\ absorption column densities of the GPS/CSO galaxies with their 
optical and radio properties, and with those of the general population of 
radio galaxies. We find that,\\
1) GPS/CSO galaxies 
show a wide range in absorption column densities with a distribution
not different from that of the general population of radio galaxies. 
We therefore find no evidence that GPS/CSO galaxies
could reside in a significantly 
more dense circumnuclear environment such that they could be ``frustrated''
radio sources $-$ hampered in their development.\\ 
2) The ratio of radio to X-ray luminosity is significantly higher for GPS/CSO
sources
than for classical radio sources. This is consistent with an evolution scenario
in which young, compact radio sources are more efficient radio emitters than 
large extended objects, at a constant accretion power.\\ 
3) Taking the X-ray luminosity of radio sources as a measure of their
ionisation power, we find that GPS/CSO sources 
are significantly underluminous in
their [OIII]$_{5007\AA}$ line luminosity, including a weak trend with age.
This is consistent with the fact that the Str\"omgren sphere should still
be expanding in these young objects. If true, this would mean that here we 
are witnessing the birth of the narrow line region of radio-loud AGN.

\end{abstract}
\begin{keywords}
galaxies: active -- X-ray: galaxies
\end{keywords}

\begin{table*}
 \centering
 \begin{minipage}{\textwidth}
  \caption{The sample of GPS/CSO radio sources with $|b|>20$\degr\ from 
the \citet{pearson88} catalogue. 
Indicated are, (column 1) the coordinates, 
(column 2) redshift, $z$, 
(column 3) radio flux density at 5~GHz, $S_{5 \rm GHz}$ (erg s$^{-1}$), 
(column 4) radio luminosity at 5~GHz, $L_{5\rm GHz}$\ (erg s$^{-1}$), 
(column 5) the 5007~\AA \oiii\ emission line luminosity,
$L_{\rm [O III]}$\ (erg s$^{-1}$) from \citet{lawrence96}, 
and (column 6) the kinematic age of the radio hot spots 
\citep[][except for B1358+624]{polatidis03}
\label{tab-sample}}
  \begin{tabular}{@{}lllcccc@{}}\hline
Source  &   Position                           & $z$ & $S_{5 \rm GHz}$\footnote{Obtained from the NASA/IPAC Extragalactic Database (NED)}& $\log L_{5\rm GHz}$ & 
$\log L_{\rm [O III]}$ & Kinematic Age\\
         &      (J2000)                                       &   &  Jy   &   & & yr\\
 \hline
B0108+388 & $01^h11^m37.3^s$\ +39\degr 06\arcmin 28\arcsec  & 0.668 & 1.6 &  44.0 &40.8 & $570\pm50$\\
B0710+439 & $07^h13^m38.1^s$\ +43\degr 49\arcmin 17\arcsec  & 0.518 & 1.6 & 43.7 & 42.4 & $930\pm100$\\
B1031+567 & $10^h35^m07.0^s$\ +56\degr 28\arcmin 47\arcsec  & 0.45  & 1.3 & 43.4 & 41.6 & $1800\pm600$\\
B1358+624 & $14^h00^m28.6^s$\ +62\degr 10\arcmin 39\arcsec  & 0.431 & 1.8 & 43.5 & 41.8 & $2400\pm1000$ 
\footnote{
The kinematic age of B1358+624 has not been directly measured, but is based on 
on its size and the average size-age
relation of GPS/CSO sources in \citet{polatidis03}.}
\\
B2352+495 & $23^h55^m09.4^s$\ +49\degr 50\arcmin 08\arcsec  & 0.238 & 1.5 & 43.0 & 41.3 & $3000\pm750$\\
\hline
\end{tabular}

\end{minipage}
\end{table*}

\section{Introduction}

Ever since their discovery, it has been speculated that those compact radio
sources that show convex-shaped radio spectra at cm wavelengths, 
may be young objects \citep{shklovsky65,blake70}. As a class, they were
named Gigahertz Peaked Spectrum (GPS) sources after their characteristic 
radio spectrum \citep[see][for a review]{odea98},
most likely caused by synchrotron self absorption \citep[e.g.]{fanti90,snellen00a}. 
High resolution Very Long Baseline Interferometry 
(VLBI) observations have shown that these sources are 
typically up to a few hundred parsec in size, often 
exhibiting jet and/or lobe structures on two opposite sides from their central 
core $-$ the reason why they are also called Compact Symmetric Objects 
\citep[CSO; eg.][]{Wilkinson94}. The most compelling evidence that GPS/CSO
are indeed young radio sources comes from VLBI monitoring observations, showing
that the bright archetypal objects in this class have
hot-spot propagation velocities of $\sim$0.1-0.2c 
\citep{owsianik98a,owsianik98b,tschager00}, 
indicating kinematic ages of $\sim$10$^{3}$ years.
This is in contrast to speculations that 
GPS/CSO galaxies are small due to confinement by a particularly dense and 
clumpy  interstellar medium (ISM) that impedes the outward propagation of 
the jets \citep{vanbreugel84,odea91}.

Note that a large fraction of GPS sources, in particular those showing a 
convex radio spectrum peaking at higher frequencies than a few GHz, 
turn out to be identified with high redshift quasars (z$\sim$2-3).
Their connection with the population of GPS/CSO galaxies at lower redshift
is not clear, and they may well be a completely separate class of object
which just also happen to exhibit a convex shaped spectrum \citep{snellen99}.
By no means it has been established that the GPS quasars 
may also represent a young stage of radio source evolution.

Here we present \xmm\ X-ray observations of a small
but complete sample of all five 
GPS sources from the \citet{pearson88} 
sample with Galactic latitude $|b| > 20$\degr\ (Table~\ref{tab-sample}).
These are among the brightest GPS/CSO galaxies in the sky.
All these sources appear to be young radio loud AGN with
four out of five sources having measured  
kinematic ages of their hot spots, indicating ages of up to $\sim3000$~yr.
No kinematic age measurement for B1358+624 exists, but based on its size and the
age-size measurements of GPS/CSO sources \citep{polatidis03}
we estimate its age to be 2400 yr.
Although the sample is limited in size,
it allows us to study the correlations between their 
X-ray, optical and radio properties, and to assess how they
compare to those of mature radio galaxies.
One expects that the X-ray luminosity is largely a manifestation of the
instantaneous accretion power of the central black hole,
whereas the radio luminosity is expected to evolve 
substantially over the life time of the source \citep[e.g.]{readhead96,snellen00a}.
Moreover, X-ray absorption measurements are an excellent means to
probe the circum nuclear density of the GPS galaxies.

\begin{figure*}
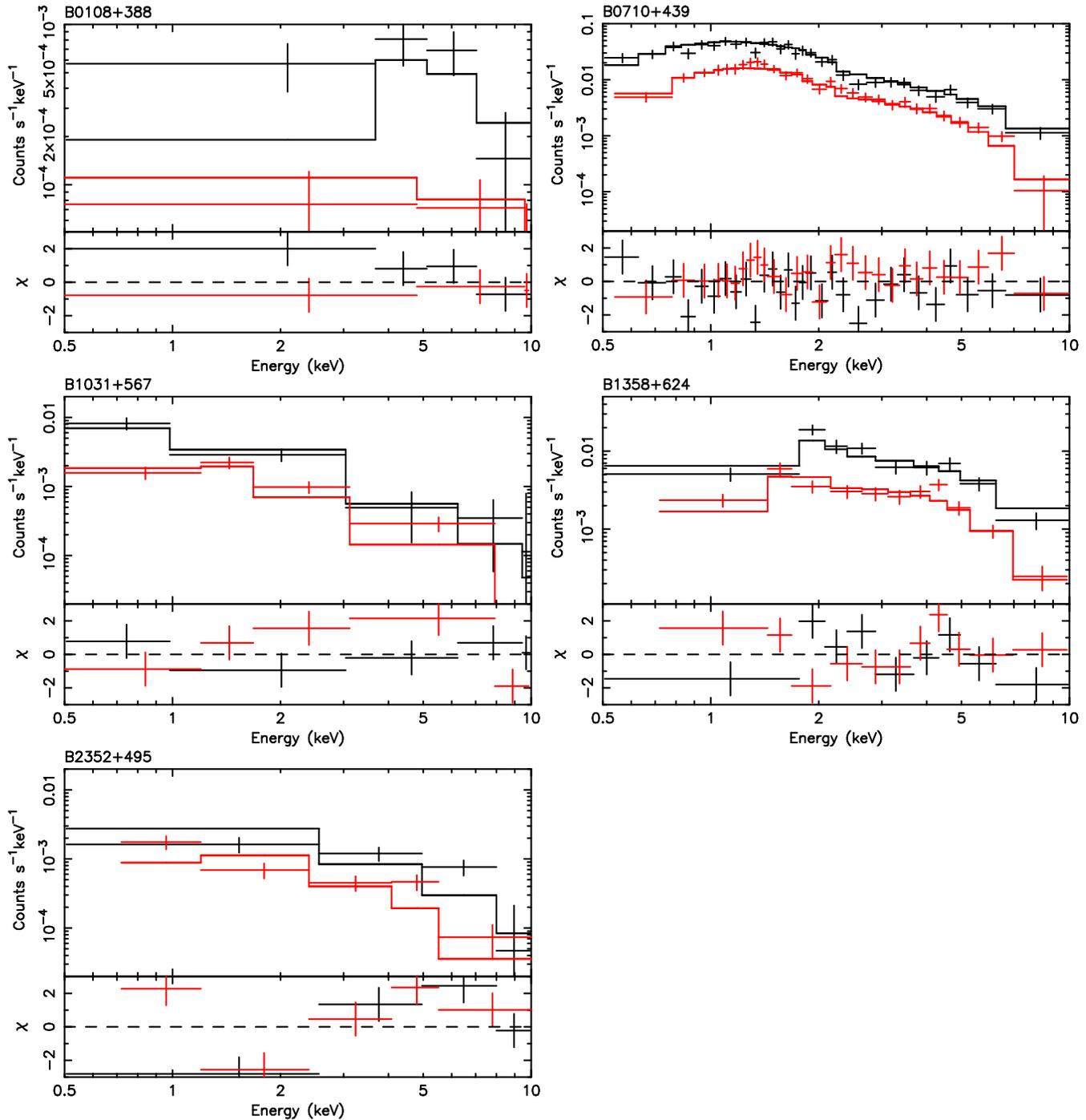

\hbox{\psfig{figure=b0108_388_ldata_delchi.ps,angle=-90,width=0.5\textwidth}
      \psfig{figure=b0710_439_ldata_delchi.ps,angle=-90,width=0.5\textwidth}
}
\hbox{\psfig{figure=b1031_567_ldata_delchi.ps,angle=-90,width=0.5\textwidth}
      \psfig{figure=b1358_624_ldata_delchi.ps,angle=-90,width=0.5\textwidth}
}
\hbox{\psfig{figure=b2352_495_ldata_delchi.ps,angle=-90,width=0.5\textwidth}
      \hskip 0.5\textwidth
}

 \caption{Observed \xmm\ count rate spectra.
The MOS spectra are shown in red and the PN spectra in black.
\label{spectra1}}
\end{figure*}

Several papers on X-ray observations of GPS sources have appeared in the 
literature, but one should be cautious to interpret these results in
terms of X-ray properties of young radio-loud AGN.
A first success was obtained by \citep{odea00} with ASCA. 
Although they did not detect B2352+495, they obtained a firm detection of 
GPS galaxy B1345+125 (PKS1345+125). 
Furthermore, \citet{guainazzi04} detected B1404+288 
(Mkn 668).\footnote{Prior to the submission of this publication we learned that
Guainazzi et al. have detected a number of other GPS/CSO galaxies in X-rays,
which do not overlap with our sample \citep{guainazzi05}}
 Both are low redshift GPS galaxies exhibiting strong optical 
line emission and are powerful infrared emitters, and may be not representative
to the class of young radio-loud AGN (yet no reliable ages have been measured
for these sources). The combined X-ray and radio observations of the
GPS {\em quasar}
B0738+313 presented by \citet{siemiginowska03b} show that, with its
prominent kpc-scale X-ray/radio jet, 
it is certainly not a young radio-loud AGN.

In order to easily compare our results for GPS/CSO galaxies with the X-ray 
properties
of a large sample of AGN by \citet{sambruna99} we adapt here a cosmology with
$H_0 = 75$~km s$^{-1}$ Mpc$^{-1}$ and $q_0 = 0.5$.

\begin{table}
 \centering
  \caption{Log of the \xmm\ observations. The MOS exposure time 
and event rates
refer to the average value for MOS1 and MOS2.\label{tab-obs}}
{
  \begin{tabular}{@{}lcccc@{}}\hline
Source  & Observation ID & Start Date & Exposures & Event rates \\
         &         &                   & (MOS/PN) & (MOS/PN)\\
         &         & d/m/y             & ks & ct s$^{-1}$\\
 \hline
B0108+388 & 0202520101  & 09/01/2004 & 16.4/12.0 & 1.6/13.6\\
B0710+439 & 0202520201  & 22/01/2004 & 14.2/11.2 & 3.2/26.6\\
B1031+567 & 0202520301  & 21/10/2004 & 22.2/12.8 & 34.4/93.0\\
B1358+624 & 0202520401  & 14/04/2004 & 12.4/12.0 & 23.2/120.6\\
B2352+495 & 0202520501  & 25/12/2004 & 15.8/12.8 & 3.7/28.5\\
\hline
\end{tabular}
}
NOTE -- The event rates refer to the total detector count rates,
not the source count rates.
\end{table}

\begin{table*}
  \begin{center}
    \begin{minipage}{\textwidth}
      \caption{Observational properties and parameters obtained by modeling the observed X-ray spectra in the range 0.5-10~keV.\label{tab-res}}
  \begin{tabular}{@{}llllll@{}}\hline\noalign{\smallskip}
                                     & B0108+388 & B0710+439    & B1031+567   & B1358+624 &B2352+495 \\
\noalign{\smallskip}\hline
\noalign{\smallskip}
PN     0.5-10~keV source count rate ($10^{-3}$~cts\,s$^{-1}$) & $4.7\pm0.9$   & $89.5\pm3.0$ &$12.6\pm2.0$ & $44.3\pm2.7$  & $8.6\pm1.2$\\
MOS1+2 0.5-10~keV source count rate ($10^{-3}$~cts\,s$^{-1}$) & $0.71\pm0.25$ & $33.0\pm1.1$ & $4.3\pm0.5$ & $15.8\pm0.9$   & $3.4\pm0.4$\\
\noalign{\smallskip}
Galactic \nh\ 
($10^{20}$cm$^{-2}$)                 & 5.80        & 8.11         & 0.56        & 1.96      &12.4\\
\noalign{\smallskip}
Normalization \footnote{Statistical errors correspond to $\Delta C=1$ (68\% confidence limits).}  
($10^{-5}$ph s$^{-1}$keV$^{-1}$ cm$^{-2}$@ 1 keV) & $3.3\pm1.1$\footnote{
The 3$\sigma$ lower limit is $1.1\times10^{-5}$ph s$^{-1}$keV$^{-1}$ cm$^{-2}$.
The source is detected at the $7\sigma$ level.} 
&$8.1\pm0.6$ & $1.2\pm0.2$ & $6.1\pm1.6$ & $1.1\pm0.2$\\
Power law slope 
\footnote{Brackets indicate that the power 
law slope was fixed to this value.}  
($\Gamma$)                           & (1.75)    & $1.59\pm0.06$ & (1.75)    & $1.24\pm0.17$ & (1.75) \\
Intrinsic \nh\ ($10^{22}$cm$^{-2} $) & $57\pm20$\footnote{
The 3$\sigma$ lower limit is $1.8\times10^{23}$~cm$^{-2}$.}
                                                 & $0.44\pm 0.08$ & $0.50\pm0.18$     & $3.0\pm0.7$ & $0.66\pm0.27$\\
\noalign{\smallskip}
Flux (2-10 keV) \footnote{Including absorption.} 
 ($10^{-13}$~erg\, s$^{-1}$cm$^{-2}$) & 0.50    &  4.0 & 0.51 & 4.8 & 0.41\\
Luminosity (2-10 keV)\footnote{Calculated for rest frame energies, ignoring absorption.}
  ($10^{44}$~erg\, s$^{-1}$)         & 1.18     & 2.16           &  0.22        &   1.67    & 0.046\\
C-statistic/bins                     &137.5/99  & 497.4/401     & 104.2/97 &    122.9/99  & 126.3/99 \\
\noalign{\smallskip}
\hline
\end{tabular}
\end{minipage}
\end{center}
\end{table*}

\begin{figure*}
\centerline{
\psfig{figure=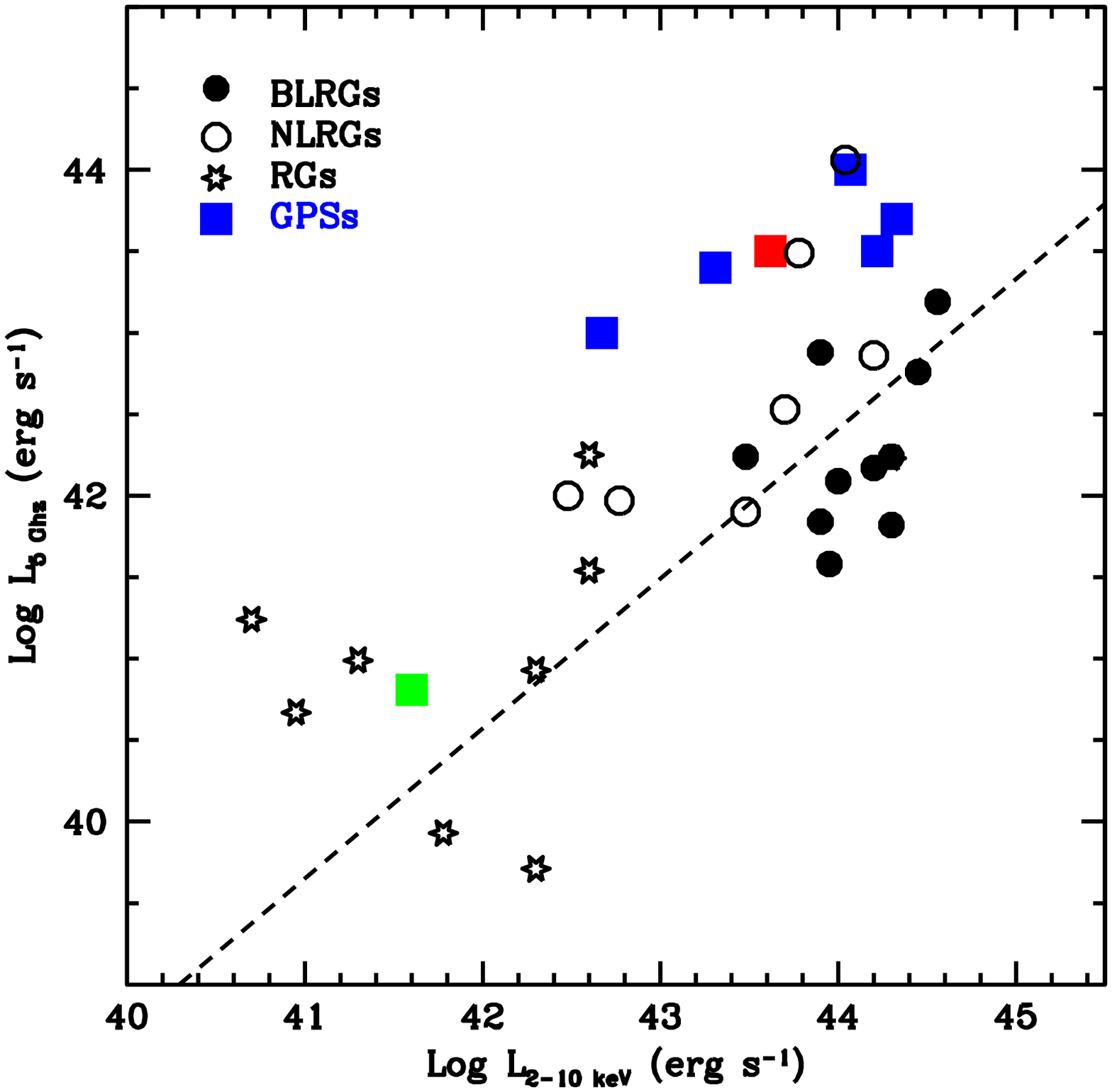,width=0.5\textwidth}
\psfig{figure=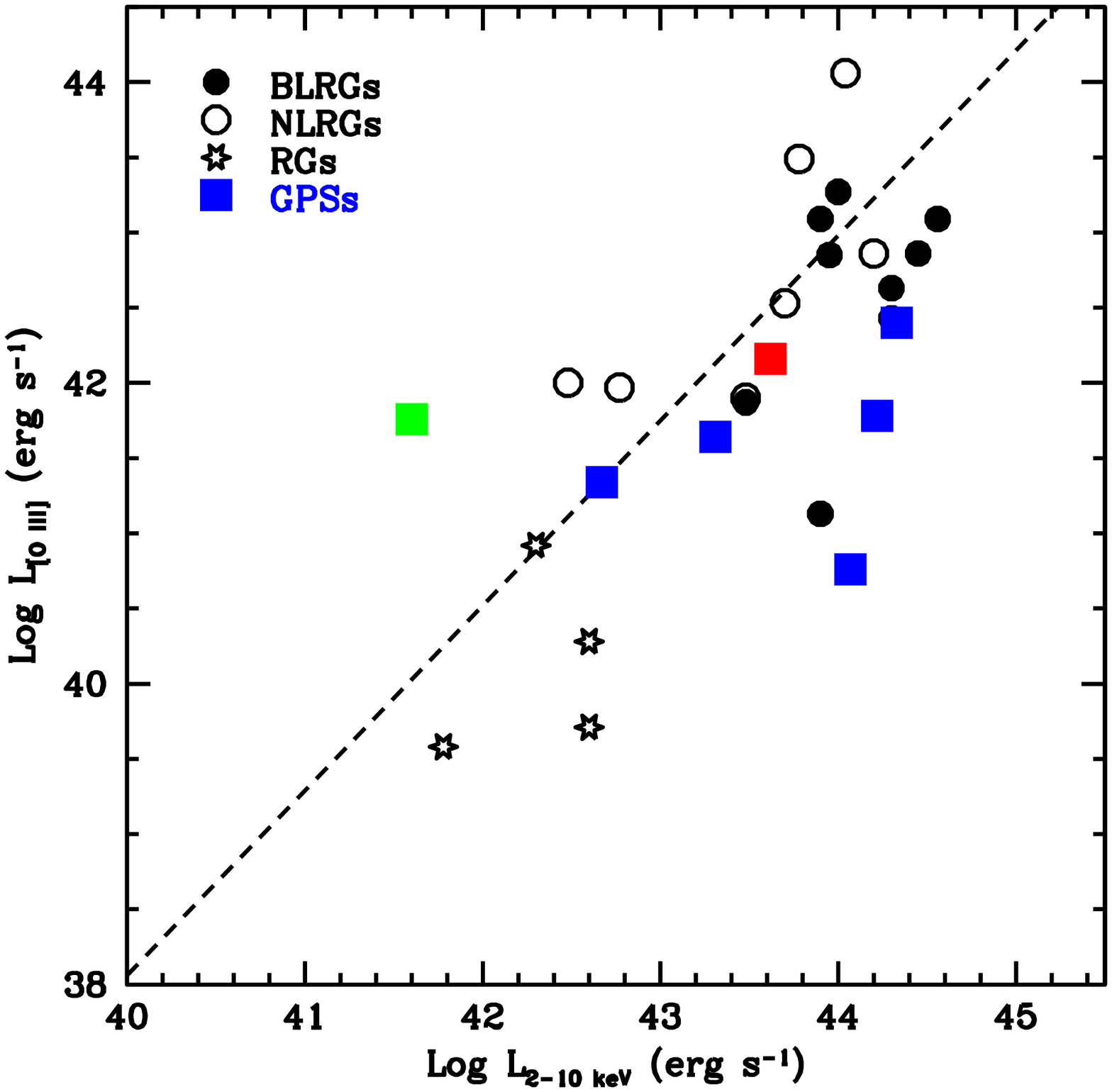,width=0.5\textwidth}
}
 \caption{
Left: Correlation between X-ray luminosity and radio luminosity 
of GPS/CSO sources (filled squares), 
compared to the radio-loud sample of active galactic nuclei
by \citet{sambruna99}.
Right: Idem, for the [O III] line emission.
In both figures the dotted lines are the average correlations for the
radio loud sample obtained by  \citet{sambruna99}.
The galaxies from the current sample are shown in blue, whereas Mkn 668
and PKS1345+125 are shown in green and red, respectively.
\label{correlations}.}
\end{figure*}

\section{Observations, spectral analysis and results}
\xmm\ \citep{jansen01} observed the five GPS/CSO sources as part of its guest 
observation program from January to December 2004 (Table~\ref{tab-obs}).
All observations were made with the  ``Thin1'' optical blocking filter.
For the data reduction we used the standard \xmm\ software package SAS v6.0.0.
Unfortunately several observations were plagued by a high particle background
(see Table~\ref{tab-obs}).
In the case of {B0108+388}, {B0710+439}, {B2325+495} 
we removed time intervals with a high background
count rate, using cut off rates of 15.5~ct\,s$^{-1}$ and 2.5~ct\,s$^{-1}$ 
for resp. PN and MOS.
For {B1031+567} and {B1358+624}
the high background persisted throughout the 
observation, and we simply used all available data.
The observation {B2325+495} had an intermediate background activity, 
so we selected time intervals with $< 40$~ct\,s$^{-1}$ and $< 5$~ct\,s$^{-1}$
for PN and MOS.
For spectral extraction we used circular extraction regions
with radii of 15\arcsec\ for {B1031+567} and {B1358+624}, and
25\arcsec\ for the other three sources.
Background spectra were obtained from rectangular regions near the source 
position, but excluding regions around 35\arcsec\ of the source.
We extracted spectra for the two MOS CCD cameras \citep{turner01}, and the PN
camera \citep{strueder01}. For each source we combined the spectra of MOS1 and 
MOS2, into one spectrum, which we analysed using averaged 
instrumental response matrices.
The potential systematic
error introduced is small compared to the statistical
errors, given the fact that the MOS1 and MOS2 are virtually 
identical instruments with similar instrumental response functions.

All the five sources of the sample are detected.
For the spectral analysis we 
employed a simple model consisting of a power law continuum
and two absorption components: One represents the 
Galactic absorption, with an absorption column, \nh, fixed
to the Galactic value of
\citet{dickey90}\footnote{We extracted the absorption
columns from  the online ``\nh-tool'', \\
\url{http://heasarc.gsfc.nasa.gov/Tools}}.
The other
absorption component corresponds to the absorption column
intrinsic to the host galaxy. The redshift of this component was fixed
to that of the galaxy, but the absorption column density was a free parameter.
The spectral analysis was done with the spectral fitting program
{\it xspec} \citep{xspec}, using the absorption models of
\citet{wilms00} (called {\it tbabs} and {\it ztbabs} in \xspec).

The slope of the power law continuum was a free parameter for
the high signal to noise spectra of {B0710+439} and {B1358+624},
but fixed to 1.75 for the other three sources whose spectra are
statistically more limited.
The best fit parameters, together with the inferred intrinsic 
X-ray luminosities between 2-10~keV are listed in 
Table~\ref{tab-res}. 
The spectra and best fit models are shown in Fig.~\ref{spectra1}.

\begin{figure}
\centerline{
\psfig{figure=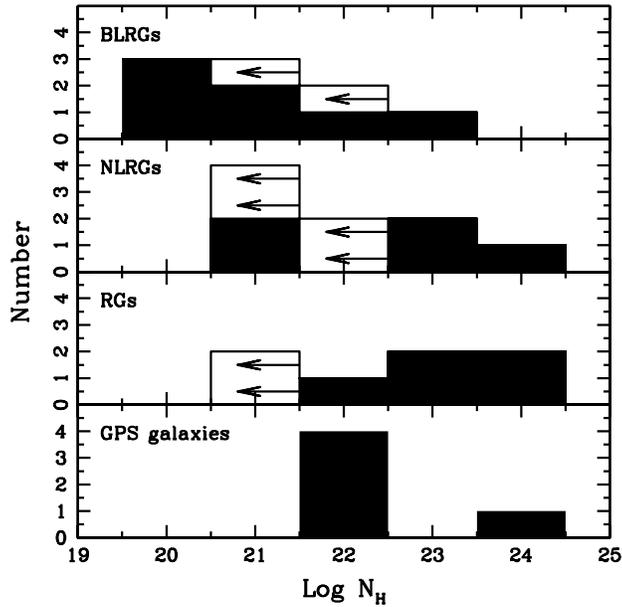,width=\columnwidth}
}
\caption{
The distribution of the intrinsic X-ray column densities for various
radio loud AGN. The lower panel are the results of the present study, the other
panels have been taken from \citet{sambruna99}.
Upper limits are indicated by arrows.
\label{fig-nhd}}
\end{figure}

\section{Interpretation}
As we are interested in whether GPS/CSO galaxies are different from
other radio-loud AGN we compare their X-ray properties
to those of the sample of radio-loud AGN whose X-ray properties were
determined by \citet{sambruna99} from ASCA observations.
We use all the sources of their broad line radio galaxies (BLRG),
narror line radio galaxies (NLRG) and radio galaxies (RG) subsamples.
Since \citet{sambruna99} do not quote upper limits for the non-detected
intrinsic X-ray asborption components, we estimate upper limits proportional
to the observed flux taking into account the errors on the detected instrinsic
absorption components.

In Fig.~\ref{correlations} we have set out the radio and \oiii\
luminosities of the galaxies in our sample to their X-ray luminosities,
and we compare them to the \citet{sambruna99} sample. 
In Fig.~\ref{fig-nhd} we show the
intrinsic absorption column density distribution.

These figures reveal three important properties of our sample of
GPS/CSO galaxies:
1) for their X-ray emission GPS/CSO galaxies are relatively radio-loud; 
2) their
\oiii\ emission is relatively low; 3) the column density distribution is
similar to those of radio-loud AGN classified by \citet{sambruna99} as 
narrow line radio galaxies (NLRGs) and radio galaxies (RGs), 
but the absorption is on average higher than those of broad line 
radio galaxies.

As we discuss below these three properties support the 
hypothesis that GPS/CSO galaxies are indeed young radio-galaxies. {
Note that there are strong indications that 
GPS radio galaxies have relatively
low [OIII] lumininosities with respect to their radio luminosities as compared
to compact steep spectrum (CSS) sources \citep{odea98}. 
This is probably the same trend
between radio, [OIII], and X-ray luminosity that we report here, 
but without the X-ray luminosity as intermediary quantity.
As noted in the
introduction, the X-ray luminosity is the quantity that
is probably the best indicator for the intrinsic power of the AGN. 
}

\begin{table}
  \caption{A comparison of the X-ray derived absorption column $\NH$
and radio absorption column measurements $N_{\rm HI}$ \citep{pihlstroem03}.
\label{tab-nh}}
\centering
  \begin{tabular}{@{}lcr@{}}\hline
&$\log \NH$ &$\log N_{\rm HI}$  \\\hline
B0108+388 & 23.8 & 21.9 \\
B1031+567  & 21.7 & $<20.1$ \\
B1358+624 & 22.5 & 20.3\\
B2352+495 & 21.8& 20.5 \\\hline
\end{tabular}
\end{table}

\begin{figure}
\centerline{
\psfig{figure=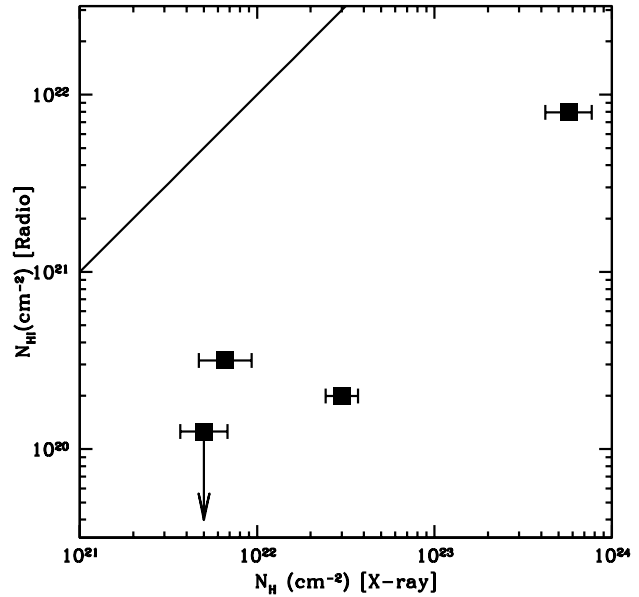,width=\columnwidth}
}
\caption{
The neutral hydrogen absorption column density, $N_{\rm HI}$, toward
the radio source \citep{pihlstroem03}
versus the intrinsic X-ray absorption column density, $\NH$, for the four
sources for which $N_{\rm HI}$\ is measured.
The solid line indicates $N_{\rm H/X-ray} = N_{\rm HI}$,
the arrow indicates an upper limit.
\label{fig-nh}}
\end{figure}

\subsection{The absorbing column density}
\label{sec-nh}
An alternative explanation for the small extent of the radio
jets in GPS/CSO galaxies that is 
still often considered in the literature \citep{odea98} is that the
radio jets are quenched by a high density in the vicinity of the nucleus,
in other words they are ``frustrated radio sources''.
It is clear from Table~\ref{tab-res} and Fig.~\ref{fig-nhd} 
that this is unlikely to be the case, as the intrinsic X-ray absorption 
is similar to other radio-loud AGN, with column densities
ranging from a relatively
modest $4\times10^{21}$~cm$^{-2}$\ to a considerably large
$6\times10^{23}$~cm$^{-2}$.
A similar conclusion was drawn by \citet{pihlstroem03} based on HI
radio absorption observations of a large sample of compact radio sources,
and by \citep{odea05} from upper limits to the molecular gas content in 
GPS sources.
Note that the X-ray absorption gives more stringent constraints on the actual
gas column densities than HI and molecular absorption densities,
as X-ray absorption depends on the total column
toward the central source, whereas radio observations only
probe the neutral fraction of the gas.
This is quite an important distinction since AGN are expected to 
create an extended ionised region, as we discuss in section~\ref{sec-oiii}.
Furthermore, X-ray absorption
probes the gas toward the accretion disk, whereas the radio absorption probes 
the neutral gas toward the radio source, which is situated at larger radii.
It is therefore not surprising that for the four galaxies in our sample
for which also HI absorption measurements have been made the HI
column is always one to two orders of magnitude lower than the X-ray
absorption column (Table~\ref{tab-nh} and Fig.~\ref{fig-nh}).
Attributing the difference solely to ionisation  effects would mean ionisation
fractions of 90\% to 99\%.
However, the ionisation fractions are likely to be lower,
because a substantial part of the X-ray absorption may
occur inside the central 100~pc, which is not probed by
absorption toward the radio hot spots.

\citet{pihlstroem03} found a strong anti-correlation between the linear
size of the radio emission and the HI column density, which they use to probe
the average density profile of the interstellar medium.
They do not consider ionisation effects, whereas this could be
an additional cause for the observed anti-correlation:
small jets are associated with young AGN, which do therefore not yet have an
extended narrow line emission region of ionised gas (see section~\ref{sec-oiii}). 
If this is the case it makes it less straightforward
to derive an average interstellar medium density profile from the
relation between $N_{\rm HI}$\ and linear size of the radio emission,
since the growth of the emission line region also depends on the
density and UV luminosity of the AGN.

\begin{figure*}
\centerline{
\psfig{figure=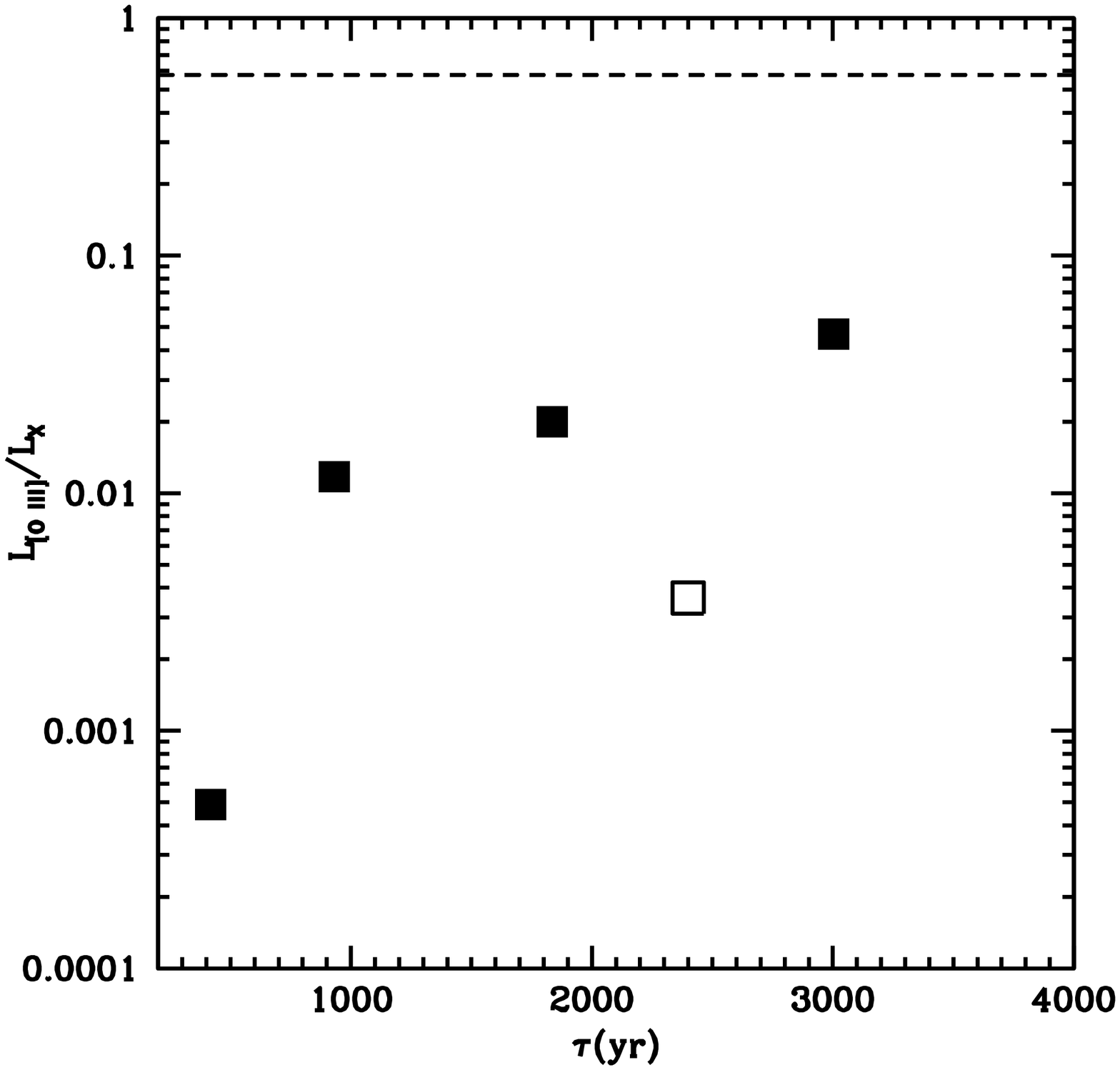,width=0.5\textwidth}
\psfig{figure=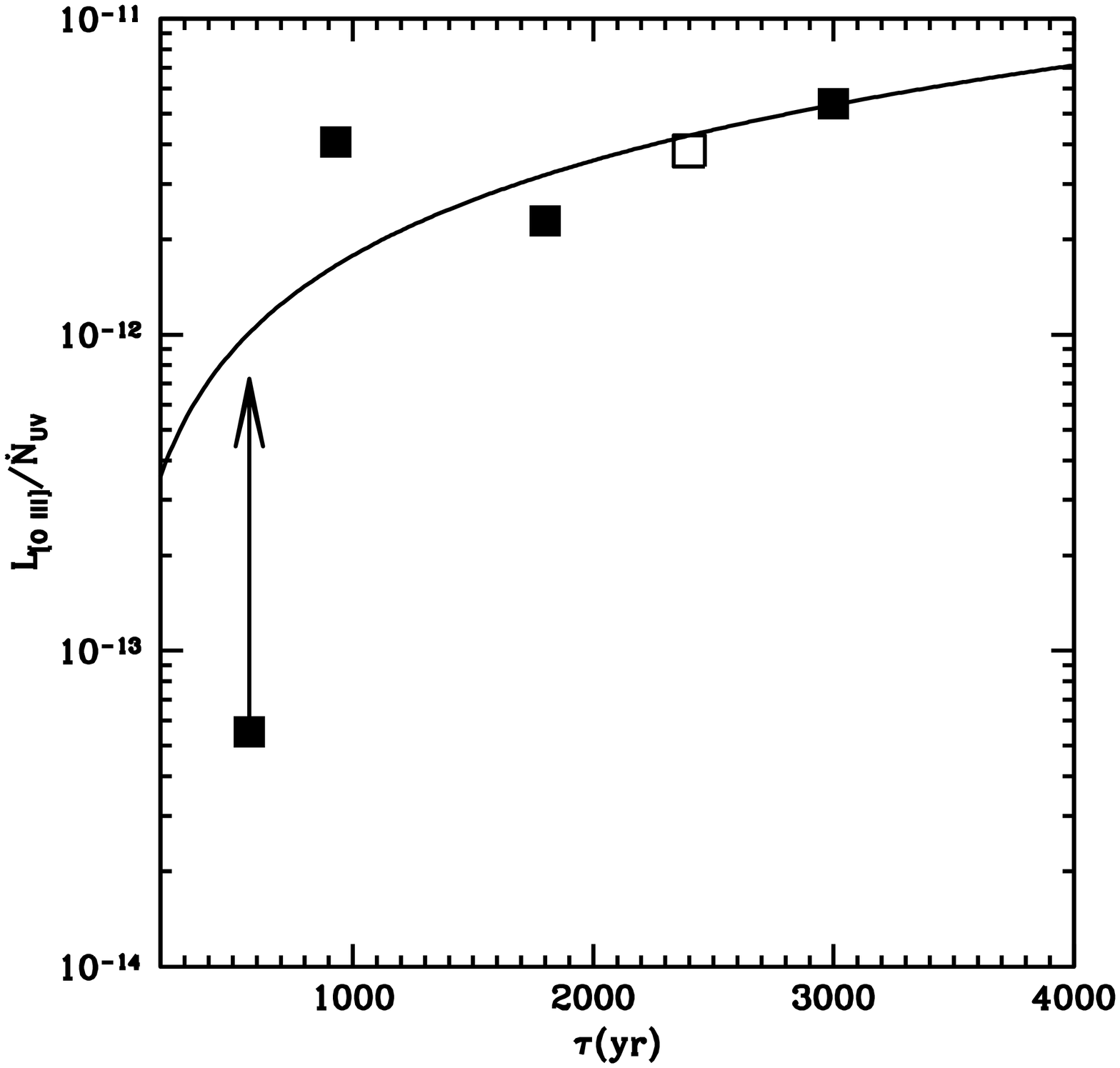,width=0.5\textwidth}
}
\caption{
Left:
The ratio of the [OIII]$_{5007 \rm\AA}$\
 and 2-10 keV X-ray luminosity versus the kinematic age of the radio source.
The dashed line
indicates  the average value taken from \citet{sambruna99}.
Right: Similar to the left hand figure, but instead of $L_X$\ the
extrapolated UV photon luminosity is used, derived from the X-ray
luminosity using the power law slope in Table~\ref{tab-res}.
The arrow indicates approximately how the ratios changes if
the UV emission is obscured by dust assuming that the dust column density
scales with $\NH$. This would only significantly affect {B0108+388}.
The line shows the expected scaling for a young expanding narrow line
region (eq.~\ref{eq-scaling}).
B1358+624 is indicated with an open symbol, as its age is
not based on a direct kinematic age measurement, but on its size 
(Table~\ref{tab-sample}). 
\label{fig-opt-age}}
\end{figure*}

\subsection{The optical line emission: 
the case for an expanding emission line region}
\label{sec-oiii}
Usually a high [O III] luminosity is taken as an indication for the
presence of a  powerful AGN, but Fig.~\ref{correlations} indicates
that GPS/CSO galaxies are relatively underluminous in [O III] compared to 
their radio or X-ray flux.
This may not be too surprising if one takes into account that
these are young radio galaxies, in which the AGN has 
switched on only a few thousand years ago.
The reason is that it takes time to establish a large emission line region
by photo-ionisation.
This is best illustrated by a simple calculation, for which we 
assume an average interstellar medium
density of 1~cm$^{-3}$\ and a typical luminosities of 
$\log L_{\rm \oiii} = 42$, and $\log L_{X} = 44$. 
The number rate of ionising photons (i.e. $> 13.6$ eV) is 
$\ndot \sim 10^{54}$ ph s$^{-1}$\ for a power law spectrum with photon
index -1.75.
Comparing this to the total number of hydrogen atoms within a 
typical region of 5~kpc radius \citep{baum89b}, one finds that
$\sim 10^{67}$ atoms have to be ionised.
In other words the central source has to shine for at least 
$\sim 10^{67}/10^{54} = 10^{13}$~s, or  $\sim$300,000 yr before it 
has completely ionized a region with a radius of 5~kpc.
This means that if an AGN has become only recently active
it must be surrounded by a small, but rapidly expanding ionisation nebula
\citep[see][for a calculation concerning the first generation of quasars]{white03}.
Hence, this would imply that in GPS/CSO galaxies we 
see the birth of the narrow line region of radio-loud AGN.

In order to see whether this is the reason that
GPS/CSO galaxies are relatively underluminous
in [O III] we consider a simplified model for the
evolution of a Str\"omgen sphere.
For an old ionisation nebula in equilibrium, the number rate
of ionising photons ($\ndot$)
should equal the number
of recombinations, from which follows the
equilibrium radius, $R_i$, of a Str\"omgen sphere:
\begin{equation}
R_i^3 = \frac{3 \ndot}{4\pi n_{\rm H} n_{\rm e}   \alpha_{\rm H}},
\end{equation}
with $\alpha_{\rm H}$\ the hydrogen recombination coefficient.
The initial rapid expansion of the ionisation nebula is described 
by e.g. \citet{spitzer68}
\begin{equation}
r_i^3 = R_i^3\{1- \exp( -n_{\rm e} \alpha_{\rm H} t)\} ,\label{ifront}
\end{equation}
with $r_i$\ the radius of the ionisation front and
$t$\ the time since the central source switched on.
For small $n_{\rm e} \alpha_{\rm H} t$ we can approximate this by
\begin{equation}
r_i^3 \approx n_{\rm e} \alpha_{\rm H} t R^3 =
  \frac{3 \ndot}{4\pi n_{\rm H} } t
\end{equation}
The \oiii\ line emissivity is then given by 
\begin{eqnarray}
\dot{N}_{\rm \oiii} = \frac{4\pi}{3} r_i^3  n_{\rm e} n_{\rm OIV} f_{5007 \rm\AA}\alpha_{\rm O III} =\nonumber\\
  n_{\rm e} \frac{n_{\rm OIV}}{ n_{\rm H}}f_{5007 \rm\AA}\ \alpha_{\rm O III} \ndot t,
\end{eqnarray}
with $f_{5007 \rm\AA}$ the probability of emitting a photon at $5007$~\AA\ after recombination.

Assuming the interstellar medium densities in the GPS/CSO galaxies are more or
less similar, we can expect the following
correlation between the \oiii\ luminosity and the number rate
of ionising photons for young GPS/CSO sources of kinematic age $\tau$, provided
that the age of the radio galaxy coincides with the
birth of the ionisation nebula:
\begin{equation}
L_{\rm \oiii} \propto \tau \ndot \label{eq-scaling}.
\end{equation}

As a first approximation we consider whether there is a relation
between the observables $L_{\rm \oiii}$, and $\tau$\ and $L_X$.
However, Fig.~\ref{fig-opt-age} illustrates that the five sources
in our sample do not support a simple
scaling of $L_{\rm \oiii} \propto \tau L_X$.
There may be various reasons why this is not the case:
e.g.
the interstellar medium density varies from galaxy to galaxy,
their is no simple proportionality between $L_X$ and $\ndot$, due
to different spectral energy distributions (different spectral slopes,
or spectral breaks), and shocks induced
by jet cloud interactions may provide an additional source of
ionisation.\footnote{Although we do not have kinematic age estimates
of {PKS1345+125} and Mkn668 \citep{guainazzi04}, reasonable values
for the age in fact show these galaxies to be too bright in \oiii\
compared to the galaxies in our sample. In this case the reason is very likely
that a large part of the \oiii\ emission is not related to the activity of the 
central nucleus, as both galaxies show evidence of recent merger activity,
and are bright infrared sources.} Moreover, one of the outliers is
B1358+624, for which we only have an approximate age.

However, a hint of what may the prime reason for deviations
from Eq.~\ref{eq-scaling} is provided by the very low \oiii\ luminosity
of {B0108+388}, because this is also the source
with the highest absorption column (Table~\ref{tab-res}). So the most
likely reason that the optical emission is lower than expected is that
the ionising UV flux is blocked by absorbing material close
to the nucleus. The absorbing material is probably not the result
of neutral hydrogen and helium, as, being close to the nucleus
it would be ionised almost immediately,
but dust grains, which
may survive the extreme conditions close to the nucleus for
1000~yr to $10^6$~yr, depending on the destruction mechanisms
and dust particle sizes \citep[e.g.][]{villar-martin01}. For a young source
like {B0108+388} this means that an appreciable amount of dust may still 
enshroud the nucleus, frustrating the formation of a narrow emission
line region.
Note that this may also explain the relatively high neutral hydrogen column
density of {B0108+388} \citep[][Fig.~\ref{fig-nh}]{pihlstroem03};
the hydrogen ionisation fraction of the inner stellar medium is likely to be 
low, which would also support the idea that photo-ionisation
is the dominant source of ionisation. If ionisation is dominated
by shocks generated by the jet, one would expect that {B0108+388} 
would be relatively bright in \oiii\ as its interstellar medium density is
apparently high. B1358+624 deviates less from the expected relation in the
right hand panel of Fig.~\ref{fig-opt-age} due
to its relatively flat X-ray spectrum, which makes that the number
flux of UV photons is relatively small with respect to the X-ray luminosity.
However, since we do not know the broad band spectral shape, we do not
want to overemphasise this.

Let us now consider the possible implications of dust absorption.
The optical depth of dust particles depends on
dust particles cross sections ($\sigma_d = \pi r^2$ with $r$
the physical size of the particles, $r\sim0.1~\mu$m) 
and the column density of dust particles $N_d$. 
To obtain an order of magnitude estimate we assume that most
dust particles consist of silicates, and that all silicon is depleted into
dust. As $N_{\rm Si} \approx 4\times10^{-4} \NH$, and
a typical dust particle density is $\rho = 3.5$~g cm$^{-3}$, we have
$N_d \approx 10^{-13} \NH$, and the optical depth should be around
$\sigma_d N_d = 3\times10^{-23} N_{\rm H}$. This means that dust particles absorb
an appreciable amount of UV flux if the hydrogen column density is comparable
to, or exceeds, $10^{23}$cm$^{-2}$, which is only the case for 
{B0108+388}. 
We have therefore extrapolated from the observed $L_X$\ the ionising
photon luminosity $\ndot$\ using the observed spectral properties.
Plotting now $\loiii$\ as a function of $\tau\ndot$ we see less scatter,
certainly if we allow for dust absorption (Fig.~\ref{fig-opt-age}).
In order to bring {B0108+388} on the expected relation
we need a conversion of $\NH$ to dust optical depth that is 16\% of the above 
order magnitude estimate. Note that in a log-log plot the absorption
enters linearly, since 
$\ln(\dot{N}_{UV-abs}) = \ln(\ndot) - \sigma_d N_d$.
Hence, only {B0108+388} 
is likely to be significantly affected by dust absorption.

The large uncertainties in extrapolating from the observed $L_X$\ to
an ionising UV flux makes that we cannot use Fig.~\ref{fig-opt-age} 
(right panel)
to prove that Eq.~\ref{eq-scaling} is an accurate description, 
but it makes it at least 
plausible that for GPS/CSO galaxies the \oiii\ emission is relatively
low due to an underdeveloped narrow emission line region. This is 
consistent with the idea that GPS-galaxies have AGN that
switched on around the same time that the radio jets were formed.

Further support for the idea that GPS/CSO galaxies are in the process of
creating an extended narrow line region comes from
the fact that neutral hydrogen apparently extends close
to the compact radio jets,
given the fact that there is strong anti-correlation between
the neutral hydrogen column density and jet-size \citep{pihlstroem03}.
Note that GPS/CSO galaxies have sizes of the order of a few 100~pc, whereas
narrow emission line regions can extend up to 10~kpc.

\begin{figure}
\centerline{
\psfig{figure=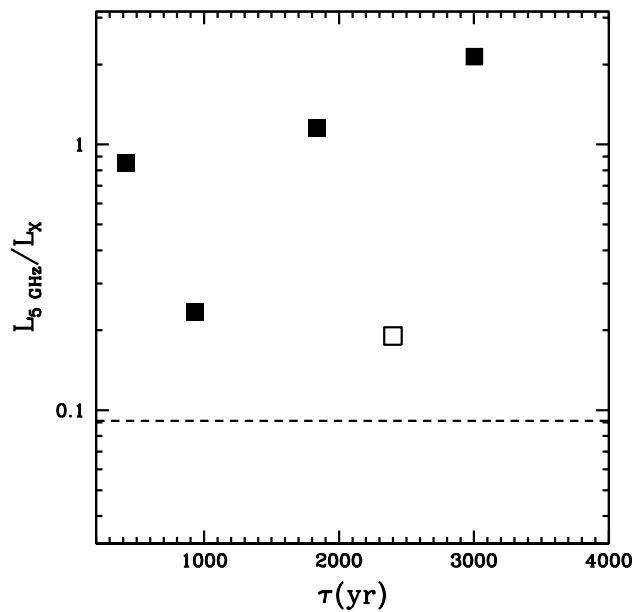,width=\columnwidth}
}
\caption{
The ratio of the radio and X-ray luminosity versus the
kinematic age of the radio source. The dashed line
shows the average ratio for
radio loud AGN galaxies with $\log L_X > 42$, taken from
\citet{sambruna99}. Only sources comparable
in X-ray luminosity to our sample were chosen, because
weaker X-ray sources appear to be relatively radio bright.
Like in Fig.~\ref{fig-opt-age}, B1358+624 is indicated with an open symbol.
\label{fig-radio-age}}
\end{figure}

\subsection{The radio luminosity versus the X-ray luminosity}
The X-ray emission from AGN is thought to come predominantly from
the immediate vicinity of the central black hole, 
i.e. thermal emission
from the accretion disk reprocessed by 
the hot plasma in its vicinity.
It is unlikely that synchrotron radiation
from the jet makes a dominant contribution to the X-ray
band. The reason is that, given the typical magnetic fields
infered from radio luminosities \citep[$>1$~mG, ][]{odea98},
the synchrotron cooling
time relevant for X-ray synchrotron radiation is in the order of
only 2~yr for an electron energy of 10~erg. This means that
a very small fraction of the total radio jet volume could
produce X-ray synchrotron emission.{
Another potential source of X-ray emission from outside the
central region could be inverse Compton emission by the 
relativistic electron population in the jets \citep[c.f.][]{belsole05}.
Although we cannot totally exclude a significant inverse Compton
contribution, it seems unlikely to be the case for our sample.
The reason is that the X-ray
emission should in that case come from the same locations as the
radio emission (the bright regions in the jets). However, we would then
expect that the measured X-ray absorption columns would be more consistent
with radio absorption measurements toward the jets, which is not the case
(section~\ref{sec-nh}).
}

It is therefore reasonable to assume
that, compared to the radio
and \oiii\ luminosity, the X-ray luminosity is more directly related to
the accretion power of the AGN.
Nevertheless, the radio and optical luminosities are still indirectly
related to the accretion power, given the
correlations between radio, optical and X-ray 
luminosities \citep[e.g.][]{sambruna99}.

It is, therefore, interesting that GPS/CSO galaxies seem on average radio
bright compared to the X-ray luminosity (Fig.~\ref{correlations}). 
Given the absorption column distribution and low \oiii\
emission, we can ignore the interpretation
that the radio emission is relatively bright because the
interstellar medium is dense in GPS/CSO galaxies, as argued by advocates
of the ``frustrated radio source'' scenario. It is, therefore,
very probable that GPS/CSO galaxies are radio bright because they are young.
However, within our sample no relation between age and radio over X-ray ratio
can be seen (Fig.~\ref{fig-radio-age}), nor is there a correlation
with column density.

Nevertheless, the fact that for the sample as a whole the radio to X-ray 
luminosity is brighter than for other radio-loud AGN 
is at least qualitatively in agreement with  several radio evolution models,
such as \citet{fanti95,readhead96,kaiser97,alexander00} and
\citet{snellen00a}.
These models describe the evolution of radio jets,
with radio brightnesses depending on the radial
density distribution of the interstellar medium. The radio jets
are relatively bright as long as the they plow through the dense
regions of the galaxies, but decline as soon as they propagate
outside the core of the galaxy.
The difference between the various models 
is that some assume a density distribution
described by a King profile \citep{snellen00a,alexander00}, 
whereas others assume a density profile
falling of as power law of radius \citep{kaiser97}.
As a result, the non-power law models predict that galaxies in the
GPS-phase are still increasing in radio luminosity in time, until the jet
has reached the core radius of $\sim 1$~kpc, after which the luminosity
declines. In this case the relatively brightest phase of
radio galaxies would be represented by the so-called compact steep spectrum 
sources (CSS), which have more extended radio jets than GPS/CSO sources. 
The power law density models predict that right after the
jet emerges from the core region the radio emission starts to decline.

Our results are inconclusive regarding the details of the early evolution
of the radio luminosity. However, future X-ray observations of CSS galaxies
could help to clarify the brightness evolution further,
as models with a King profile predict that 
CSS galaxies should, on average, have  a higher radio to
X-ray luminosity ratio than GPS/CSO galaxies, whereas models that assume
power law density profiles predict that CSS galaxies should have a smaller
radio to X-ray luminosity.

\section{Conclusions}

We have presented \xmm\ observations of a sample of all the
GPS/CSO galaxies with $|b|>20$\degr\
from the \citet{pearson88} catalog, four of which 
have measured kinematic time scales for the jet expansion.
All five of the sources are detected by \xmm\
thereby increasing the number of X-ray detected GPS/CSO galaxies
from 2 to 7.
These detections allow us to compare the X-ray properties of GPS/CSO 
galaxies with those of other radio loud AGN.

The results presented here support the hypothesis that GPS/CSO galaxies
represent the young phases in the evolution of radio-loud AGN.
The alternative explanation that the radio sources are compact due
to confinement by an exceptionally high density of the interstellar medium in 
those  galaxies, seems extremely unlikely in view of the low intrinsic
X-ray absorption column densities, 
which ranges from 
$\NH = 4\times10^{21}$~cm$^{-2}$\ to $6\times10^{23}$~cm$^{-2}$,
and has a distribution similar to other radio loud AGN.
{
After submission of our manuscript, a preprint by \citet{guainazzi05}
arrived at apparently different conclusions based on a sample
of five different GPS galaxies observed by \chandra\ and \xmm.
Only one of the five GPS galaxies in their sample 
has \nh $< 10^{22}$~cm$^{-2}$,
whereas this is 75\%$\pm$26\% for their control sample. Note, however,
that in our sample three out of five have \nh $< 10^{22}$~cm$^{-2}$.
Taking both samples together, this means that four out of ten GPS galaxies,
or 40\%$\pm$20\% have \nh $< 10^{22}$~cm$^{-2}$\ consistent with the
control sample.

The fact that the absorption columns toward GPS galaxies
are consistent with those of other radio galaxies 
suggests that the interstellar medium densities in 
the cores of GPS/CSO galaxies are similar to those of other radio loud AGN.}
A  similar conclusion was reached by \citet{pihlstroem03} based on HI radio
absorption observations, but the X-ray data provide stronger constraints, 
as the total column density 
contributes to the absorption, including ionized regions 
of the interstellar medium.
This may contribute to the fact that in all cases the X-ray column densities
are higher than the HI column densities.  A difference
between the radio column densities and the X-ray column densities is also that
the X-ray column is measured toward the central source, whereas the HI column 
density is toward the jets, which extend outside the central region.
If the difference between radio and X-ray column density is dominated
by those geometrical effects, this would be additional evidence against the
confinement scenario, since the jets have apparently been able to pierce
through the dense local regions that contribute most to the X-ray absorption
column.

Although the $\NH$\ distribution of our sample cannot be distinguished from 
other radio-loud AGN, the ratio of radio to X-ray luminosity shows that GPS 
galaxies have a strong tendency to be relatively radio bright. This supports 
the view that for the same thrust of the jets
younger radio sources are relatively bright \citep{kaiser97,snellen00a}. 
However, the data is inconclusive concerning whether GPS-galaxies 
represent the most radio-luminous phases in the
lifes of radio-loud AGN, as would be the case if the source develops in a 
density profile
that drops of as power law with distance \citep{kaiser97,pihlstroem03}, 
or whether they are still
in their brightening phase. This would be the case
if the interstellar medium is best described by a King profile with
a relative uniform density within $\sim$1~kpc of the center and then dropping 
of as a power law \citep[e.g.][]{snellen00a}. 
In the latter case the brightest evolutionary phase of  radio-loud AGN would
be represented by the compact steep spectrum sources (CSS), which have  more 
extended radio emission than GPS/CSO galaxies. A similar study to this one 
concerning CSS-galaxies can clarify this issue.

Finally, we find that GPS/CSO galaxies are relatively weak in \oiii\ line 
emission. This is again in
support of the idea that GPS/CSO galaxies represent the very earliest stages 
of 
the evolution of radio-loud AGN, since narrow line regions need time to build 
up to their equilibrium size, and young narrow line regions are therefore not 
as bright as fully developed ones.
The narrow line region is powered by the UV flux of the central source, 
but also shocks
induced by the expanding jets are likely to contribute to their formation. 
For those very
young radio-loud AGN we advocate here that their
emission line regions are powered by 
the UV flux from the central sources. 
A case in point is that the relatively weakest \oiii\
source, {B0108+388}, has also the highest X-ray column density, 
which suggests that a dusty torus is partially blocking the UV light.
The low \oiii\ luminosity therefore shows that the birth of the narrow
emission line region must coincide more or less with the birth of the radio
jet. {
However, we caution that we only have a limited knowledge
of the nature of the ionization mechanism for the [O III] line emission.
i.e. both shocks from the jets, as the UV radiation from the
AGN may contribute to the ionization. Moreover, compact steep spectrum (CSS)
sources,
probably representing a more advanced evolutionary state of radio galaxies
than GPS galaxies, show that the forbidden line emission tends to be 
aligned with the radio jet \citep{devries99}.
This phenomenon is not well understood, but it should be accounted
for if one wants to build  a more detailed model of the evolution of 
emission line nebulae in radio galaxies.}

In summary, the findings from this X-ray study lends further support
to the theory that GPS/CSO galaxies represent the early phases of radio-loud 
AGN,
in which also the narrow line emission nebula is still in the early
phases of its evolution. Future X-ray studies may help to further clarify
the relation between age or jet-size, the extent or brightness of the emission
line region and the power of the X-ray emission. It is important
to include also compact steep spectrum (CSS) sources in such a study, as they
are likely to represent the next phase in the evolution of radio-loud AGN.
Comparing their X-ray to radio luminosity may help clarify
whether the radio emission declines already during the GPS-phase, 
or first increases, then peaks around the CSS-phase and from then on weakens.

\section*{Acknowledgments}
We thank for Elisa Costantini for helpful discussions on dust 
grain properties.
The Space Research Organization of the Netherlands is 
supported financially by NWO, 
the Netherlands Organization for Scientific Research.
This research has made use of the NASA/IPAC Extragalactic Database (NED)
which is operated by the Jet Propulsion Laboratory, California Institute
of Technology, under contract with the National Aeronautics and Space
Administration.
\xmm\ is an ESA science mission, with instruments and
contributions directly funded by the ESA member states and the USA (NASA).


\end{document}